\documentstyle[twocolumn,tighten,prl,aps,epsf,floats]{revtex}

\begin{document}
\setlength{\textfloatsep}{0pt}

\wideabs{

\title{Ordered clusters and dynamical states of particles in a vibrated fluid}

\author{
Greg A. Voth$^1$, B. Bigger$^1$, M.R. Buckley$^1$, W.
Losert$^{1,*}$ , M.P. Brenner$^2$, H.A. Stone$^2$, and J.P.
Gollub$^{1,3}$}

\address{$^{1}$Department of Physics, Haverford College, Haverford
PA 19041, U.S.A.\\
$^{2}$Division of Applied Sciences, Harvard University, Cambridge
MA 02138, U.S.A \\
$^{3}$Department of Physics, University of Pennsylvania,
Philadelphia PA 19104, U.S.A \\
$^*$Present address: IREAP, University of Maryland, College Park,
MD 20742-3511}

\date{\today}

\maketitle

\begin{abstract}

Fluid-mediated interactions between particles in a vibrating fluid
lead to both long range attraction and short range repulsion. The
resulting patterns include hexagonally ordered micro-crystallites,
time-periodic structures, and chaotic fluctuating patterns with
complex dynamics. A model based on streaming flow gives a good
quantitative account of the attractive part of the interaction.

PACS: 45.70.Qj,47.15.Cb, 05.65.+b
\end{abstract}


} 


\newpage

Many important systems consist of granular material flowing while
immersed in fluid.  In early experiments, Bagnold used fluid
saturated granular flows to remove the effects of
gravity~\cite{Bagnold1954}. Others have avoided interstitial fluid
(even removing air) in order to observe pure granular behavior.
However, the fluid can produce interactions between particles
resulting in interesting macroscopic effects.  For example,
granular heaping due to fluid interactions has been investigated
in vertically vibrated deep granular
beds~\cite{Pak1995,Smith2001}. Also, a variety of interaction
effects are important in sedimentation (for
example~\cite{Joseph1987}). Several systems show combinations of
attractive and repulsive interactions which result in ordering of
particles~\cite{Trau1996,Yeh1997,Gong2001,Grzybowski2000}.

In this letter, we present experimental studies of novel phenomena
that occur as a consequence of  both attractive and repulsive
interactions between non-brownian particles when they are vibrated
in a fluid. The observations include clustering, ordered
crystalline patterns, and dynamical fluctuating states. The fluid
mediated interactions can be tuned to produce a striking variety
of dynamical phenomena, and this provides an interesting new
mechanism for self-assembly of ordered structures.

The attractive interaction can be understood quantitatively using
a theory based on the mean streaming flow generated by the
oscillating particles.  A short-range repulsive interaction is
also observed at large vibrational acceleration. This repulsion
shows a very sharp onset as the acceleration is increased.
Together the combination of attraction and repulsion results in
particles being bound together without contact over a range of
parameters. We demonstrate how these interactions allow
micro-crystallites to form, for example hexagons surrounding a
particle center. Small numbers of particles can form stable
structures, while larger numbers of particles ($>7$) move
chaotically in a bound state with no long range order, a
``mesoscopic liquid''.

\begin{figure}[b]
\epsfxsize=3.1in
\begin{center}
\centerline{ \epsffile{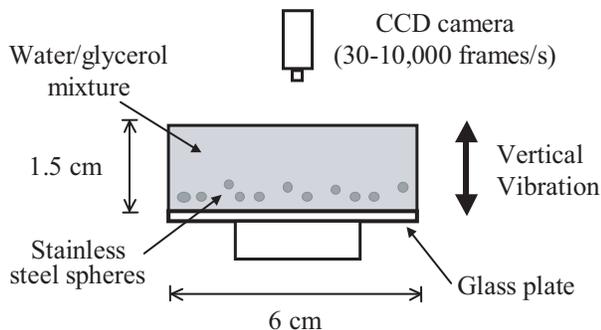} } \caption{Schematic of the
apparatus in which particles are vibrated vertically in a viscous
fluid. The vertical height of the cell is 19 times the particle
diameter, so it approximates particles bouncing in a semi-infinite
domain.} \label{fig:apparatus}
\end{center}
\end{figure}

The experimental setup in shown in Fig. 1.  We conduct all
experiments in a rigid 6 cm diameter by 1.5 cm tall cylindrical
aluminum container vibrated vertically by an electromagnetic
vibrator. We completely fill the container with a water/glycerol
mixture of kinematic viscosity $\nu = 8$ cS (density=1.15
gm/cm$^3$) and a sub-monolayer of uniform stainless steel spheres
of diameter $d=0.794$~mm and density 8.0 gm/cm$^3$. A glass window
seals the top of the cylinder to provide optical access and to
avoid surface waves. The bottom plate is made of glass since its
smoothness reduces random horizontal particle motion. We have used
both flat glass and a concave lens for the bottom surface. The
concave lens (focal length -1 m) makes no measurable change in the
particle interactions; it is used to keep particles from slowly
drifting to the edges of the cell due to imperfect leveling or
slightly non-linear vibrational motion.

Both the frequency ($\omega=2 \pi f$) and amplitude ($M$) of the
vibration of the container are controlled externally. The
nondimensional acceleration is given by $\Gamma=\omega^2M/g$.  The
container oscillations cause the particles to vibrate vertically,
typically contacting the cell bottom once each cycle. The
particles remain within 1 or 2 particle diameters from the bottom
over a wide range of accelerations.    We image the entire system
from above using a fast CCD camera.  When the system is
illuminated from an oblique angle, it is possible to measure the
positions of the shadows of particles, and thereby to determine
the amplitude of their vertical motion, $A$, defined as half the
peak to peak displacement.  The particle Reynolds number, $Re=d A
\omega/\nu$, ranges from 2 to 10.

\begin{figure}[tb]
\epsfxsize=2.5in
\begin{center}
\centerline{ \epsffile{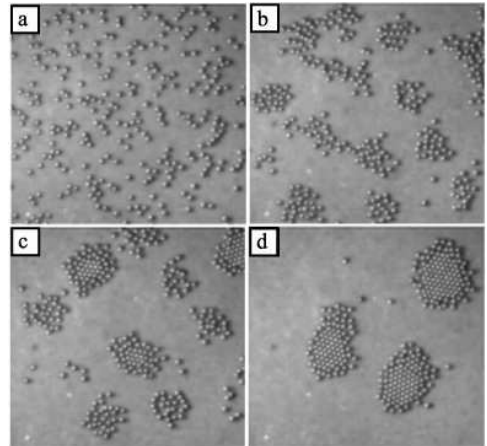} } \caption{Time evolution of an
initially random distribution of beads.  The attraction created by
the streaming flow quickly collects the particles into clusters.
($f$=50 Hz, $\Gamma$=4.5) a) t=0 s; b) t=8 s, c) t=16 s, d) t=32
s.}\label{fig:clustering}
\end{center}\end{figure}

In Fig. 2 we show the time evolution of an initially random
distribution of particles.  After the vibrator is started, the
particles quickly collect into localized clusters, and the
clusters then slowly coalesce in a manner reminiscent of
coarsening in phase transitions.

A two particle system provides a particularly simple flow in which
the particle attraction can be compared with theory.  Figure 3
shows the distance between the centers of two particles as a
function of time.  The approach rate increases until they come
into contact.  The shape of these curves provides clear evidence
that the clustering is due to a fluid-mediated interaction, and
not to inelastic collisions, which cause clustering in a similar
system with no fluid~\cite{olafsen1998}. The separation also shows
a small periodic modulation with the period of the vertical
vibration. Particles separate near their lowest point and approach
near the highest point of their motion.

The attractive interaction can be quantified by considering the
flow produced by each individual particle's motion. The vibrations
produce an oscillatory flow around each particle, which is well
approximated as a potential flow, outside of a boundary layer with
approximate thickness $\delta_{osc}=\sqrt{2\nu/\omega}$. In the
experiment at $f=50$ Hz, $\delta_{osc} =0.2~{\rm mm}$, which is
smaller than the particle radius, $a=0.397$~mm. The potential flow
near each oscillating particle is ${\bf u}=\nabla \phi$, where
$\phi=-\frac 12 A \omega \sin(\omega t) a^3 \cos(\theta)/r^2$;
here the origin is at the middle of the particle, and $\theta$ is
the angle between a given location on the sphere and the forcing
direction. At the particle surface, the component of the velocity
field parallel to its surface is $u_{||}(\theta)= \frac 12 A
\omega sin(\omega t) sin(\theta)$~\cite{images}.

Long ago Lord Rayleigh pointed out that when the magnitude of the
oscillatory flow $u_{||}=u_{||}(\theta)$ varies along a solid
surface, such as that of the particle, a {\it steady} secondary
flow is generated~\cite{Rayleigh1883,Riley2001}. For flow near the
surface, mass conservation then requires that there is also a flow
perpendicular to the boundary with magnitude $u_\perp \sim
\delta_{osc} \partial_{||} u_{||}$. Since this flow is not
entirely out of phase with $u_{||}$ in the boundary layer, every
oscillation cycle transports a finite amount of momentum into the
boundary layer. Hence there is a time independent force density on
the fluid in the boundary layer, which is parallel to the boundary
and of order $\rho \langle u_{||}\partial_{||} u_{||} \rangle$,
where $\rho$ is the fluid density and $\langle \cdot \rangle$
denotes time average over a cycle. This forcing produces a steady
flow, which when balanced against the viscous force $\rho \nu
u_{steady}/\delta_{osc}^2$ gives the magnitude of the steady flow
at the edge of the oscillatory boundary layer $u_{steady} \sim
u_{||}\partial_{||} u_{||}/\omega$.

Hence,  a steady flow is produced with a magnitude $u_{steady}
\sim A^2 \omega/a$. A careful analysis \cite{Amin1990} shows that
this steady flow pushes fluid away from the poles of each
particle; and so there is a perpendicular inflow velocity towards
the equator. This steady inflow is the origin of the attractive
interactions between two particles.

To compute the rate at which particles come together, we must
determine the inflow velocity far from the particle.  Besides the
oscillatory boundary layer, there is also a boundary layer caused
by the steady flow itself~\cite{Stuart}, with a scale
$\delta_{steady}=a/\sqrt{\rm Re_{steady}}$, where ${\rm
Re_{steady}}= A^2 \omega/\nu$ is the Reynolds number of the steady
secondary flow.  A matching argument, connecting the flows in the
boundary layers to a potential flow far away, determines the
inflow velocity to be $v(r)=-0.53 A \sqrt{\omega \nu} a^2/r^3$ at
a distance $r$ from the center of the sphere. Thus, if $R(t)$
denotes the distance between the particle centers, it follows that
$dR/dt=2 v(R)$ since the particles follow the slow changes in the
horizontal motion of the fluid.  This implies that the separation
between the two spheres should decrease according to the law
\begin{equation}
R(t) =(R_0^4-4.24 A \sqrt{\omega\nu} a^2 t)^{1/4}.
\label{eq:approach}
\end{equation}
This formula assumes that (i) $\delta_{osc}/\delta_{steady}<1$,
which can be rewritten $A < a/\sqrt{2}$; (ii) the particles are
far enough apart that they do not affect each other's boundary
layers; and (iii) neglects the influence of the bottom plate. Our
estimates indicate that assumption (i) holds when $\Gamma$ is
small or $f$ is large; (ii) holds except at the final moments of
approach; and (iii) holds except during the small interval in each
cycle when the particle bounces off the plate.

\begin{figure}[tb]
\epsfxsize=2.7in
\begin{center}
\centerline{ \epsffile{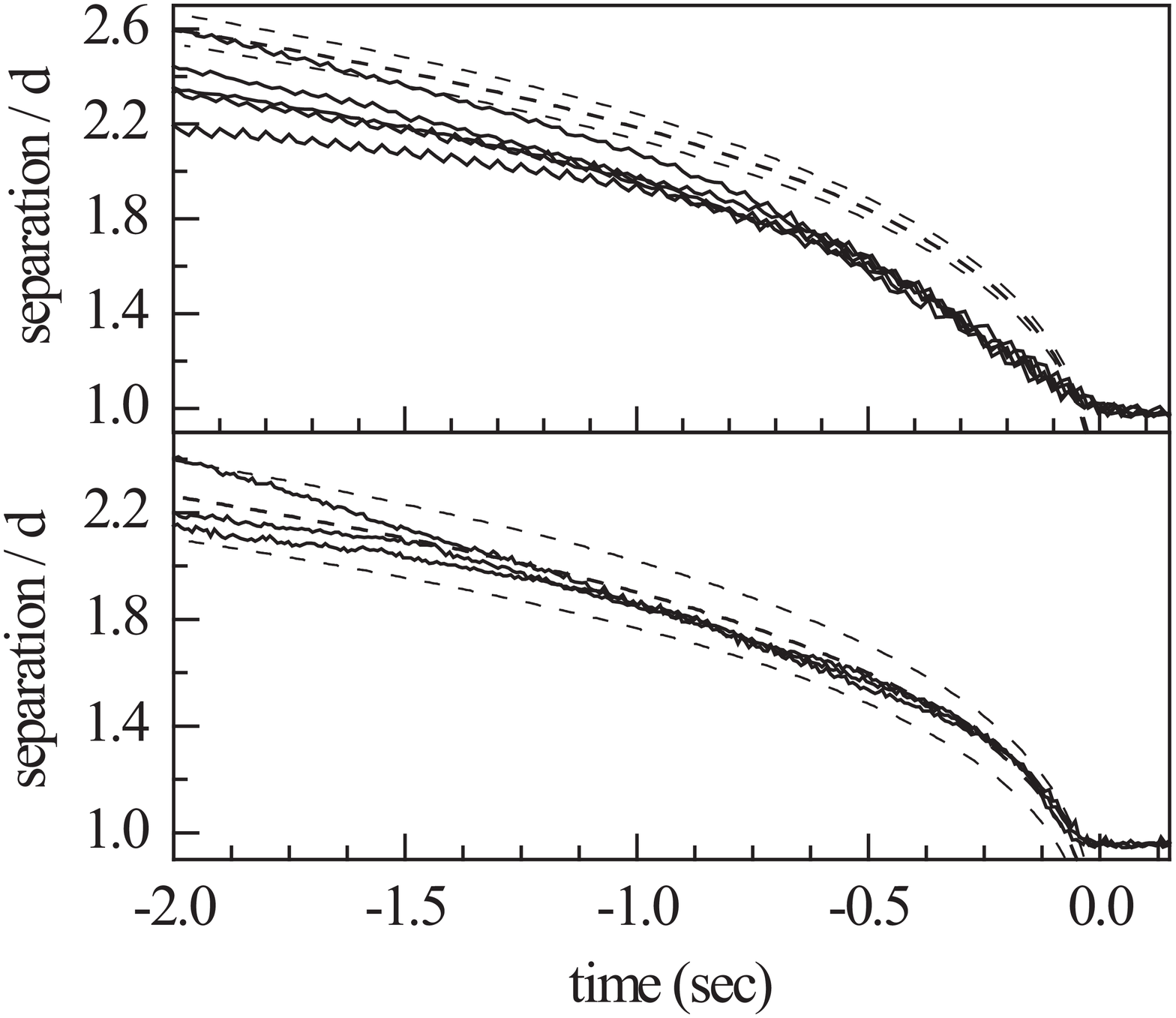} } \caption{Distance between two
particles as they are brought together by the attractive flow. The
solid curves are different experimental runs, and the central
dashed curve shows the approach predicted by the steady streaming
theory.  Upper and lower dashed curves show the effect of
measurement uncertainty in $A$.  The time axes have been shifted
so that the separation extrapolates to zero at $t=0$. (a) $f$=20
Hz, $\Gamma=2.9$, $A=0.41 \pm 0.04$~mm. Here the vertical particle
motion is periodic. (b) $f$=50 Hz, $\Gamma=4.6$, $A=0.15 \pm
0.04$~mm. The vertical motion is now chaotic and the mean
oscillation amplitude is used for $A$ in plotting the theoretical
curve.} \label{fig:approach}
\end{center}
\end{figure}

In Fig. 3, the central dashed line shows the approach curves
predicted by Eq.~\ref{eq:approach} for the conditions of the
experiment. The agreement of the data with both the predicted
functional form and rate of approach is quite good considering
that all the parameters used by the theory are independently
determined. At 20 Hz (Fig. 3a), the measured trajectories approach
somewhat more slowly than the theory predicts. At 50 Hz (Fig 3b),
the measured approach rate is closer to the prediction, and the
functional form agrees better, most likely because the oscillatory
boundary layer is thinner in this case and so assumption (i) above
is better satisfied. The upper and lower dashed curves show the
effect of measurment uncertainty in $A$. Since at 50 Hz the
amplitude is very small and the vertical motion is chaotic, the
uncertainty is rather large in this case. Overall, the theory
quite accurately captures the approach of two particles, and it
seems unambiguous that the attraction is caused by the streaming
mechanism.

\begin{figure}[tb]
\epsfxsize=2.7in
\begin{center}
\centerline{ \epsffile{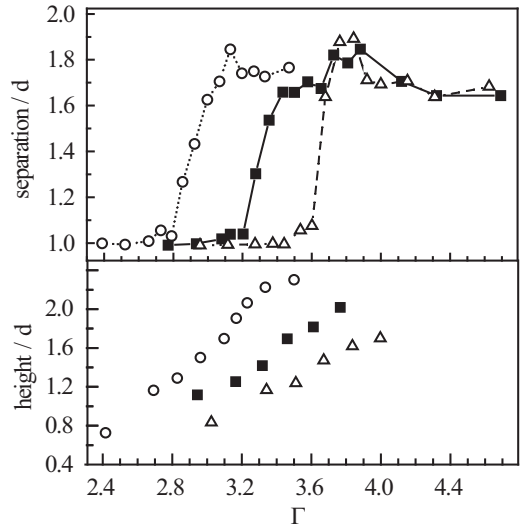} } \vspace{5pt} \caption{(a) For
two particles, the equilibrium distance between particle centers
as a function of acceleration shows a rapid transition from
contact to a nonzero separation. Separation is normalized by the
particle diameter, $d$. (b) Measurements of the height that
particles bounce above the plate show no dramatic change at the
transition. Frequencies are 17 Hz (open circles), 20 Hz (filled
squares), and 23 Hz (open triangles). Larger frequencies have the
transition at larger acceleration.} \label{fig:transition}
\end{center}
\end{figure}

At larger accelerations, a short-range repulsion becomes prominent
and limits the closeness of approach. This can be seen in Fig. 4,
which displays the steady state separation between two particles
as a function of $\Gamma$ for several different forcing
frequencies. At each frequency, there is a step-like rise in
separation at a characteristic $\Gamma$ that moves to higher
$\Gamma$ as the frequency increases. It is interesting to note
(Figure 4b) that there is no sharp change in the amplitude of the
vertical motion of the particles with increasing $\Gamma$, so the
onset of non-zero separation must be due to the fluid flow and not
to changes in the bouncing dynamics.  The onset of separation at
each frequency appears when the particles have a peak-to-peak
vertical amplitude of approximately $1.4d$.

A fully quantitative theory of this effect is not available. One
possible explanation is that the repulsion becomes important when
the oscillatory boundary layer is thicker than the steady boundary
layer.  An alternate possibility suggested by flow visualization
is that the repulsion is due to recirculating zones near the
particles created by deflection of the downward part of the
streaming flow by the plate. Explaining the details of the
repulsive interaction remains as a challenge for future work.

\begin{figure}[tb]
\epsfxsize=2.7in
\begin{center}
\centerline{ \epsffile{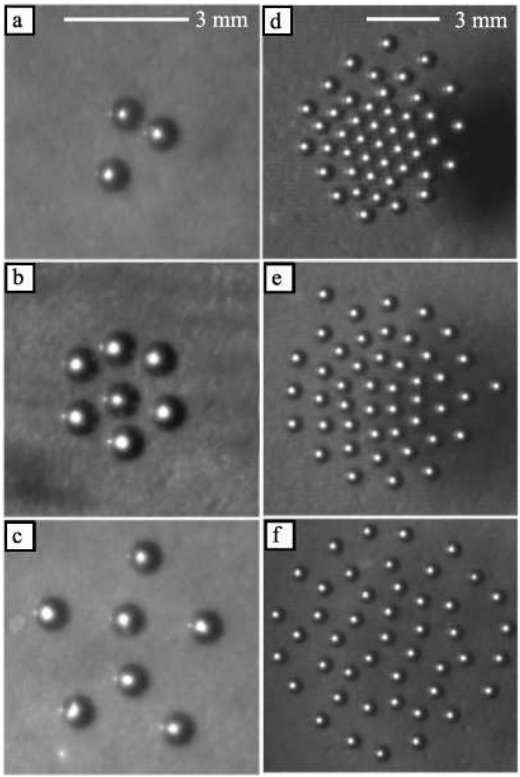} } \vspace{5pt} \caption{Patterns
formed by multiple particle systems when both attraction and
repulsion are important.  All images are acquired at f=20 Hz. (a)
3 particles at $\Gamma=3.0$. (b) 7 particles in a stable hexagon
at $\Gamma=3.0$. (c) 7 particles in stable time-periodic motion at
$\Gamma=3.7$. (d) Many particles at $\Gamma=3.7$, where the
central particles are in contact and exterior particles are held
apart by the hydrodynamic repulsion. (e) At $\Gamma=3.9$ all the
particles have separated and form a bound liquid. (f)
$\Gamma=5.3$, apparently chaotic. The time-dependent character of
states (c) through (f) are revealed in on-line
animations~{\protect \cite{animations}}.} \label{fig:patterns}
\end{center}
\end{figure}

Systems of more than two particles display a range of fascinating
patterns and dynamics, including ordered crystallites,
time-dependent ordered patterns, and bound states with complex
particle motion. For example, one might expect that when
attraction and repulsion are both important, the three particle
system would form a stable symmetric triangular configuration. For
low acceleration this is nearly the case, but as the acceleration
increases, the system breaks symmetry to a state with two paired
and one distant particle as shown in Fig. 5(a). This state is
fairly stable, but over long times the distant particle can wander
close enough to cause a new pairing.

In a system with 7 particles, stable hexagonal states form over
the range $2.8 < \Gamma < 3.0$ (Fig. 5b). At slightly higher
acceleration, the system switches to a time dependent state with 2
central particles rotating inside a ring of 5 others (Fig 5c).
Particle motion here is periodic and quite stable, and an
animation of this ``dance'' is available online~\cite{animations}.
At yet higher acceleration, 7 particle clusters exhibit apparently
chaotic motion.

For larger numbers of particles, the presence of both attraction
and repulsion leads to even more complex many body effects, which
are generally time-dependent. Figures 5(d-f) show a system of many
particles at 3 different accelerations.  Here the onset of a
non-zero separation between particles occurs at higher
acceleration than for the two particle case.  As shown in Fig.
5(d) ($\Gamma=3.7$), particles near the center of the cluster can
remain in contact even when particles near the periphery have
separated. One reason for this is that the vertical motion of
particles is significantly smaller at the center of the cluster
than at the edges, causing weaker repulsion. At $\Gamma=3.9$, (Fig
5e) the inner particles have separated and the system forms a
bound state with weakly chaotic motion of all the particles, a
``mesoscopic liquid''. One interesting feature of this state is
that the preferred distance between particles results in the
formation of closed shells.  At the parameters of Fig 5(e) there
are two more particles than fit in the closed shells, and these
``valence'' particles move more freely than the others. Further
increase of $\Gamma$ results in increased interparticle distance
and more rapid particle motion (Fig 5f). Animations of these
time-dependent states are available~\cite{animations}.

We emphasize that the chaotic motion of particles in the clusters
is not a result of external effects such as roughness of the
driving surface.  A single particle bounces regularly with
possibly a slow drift due to imperfect leveling or slightly
non-vertical vibrator motion.  Multiple particles exhibit
complicated motion as a result of the nonlinearity of the flow of
the interstitial fluid.

In summary, we have shown that fluid mediated interactions between
particles can be delicately tuned to produce a great variety of
ordered and time-dependent dynamical states of these many body
systems.

This work has been supported in part by NSF grants DMR-0079909 to
Haverford College, DMR-0072203 to The University of Pennsylvania,
and DMS-0296056 to Harvard Universtiy. We appreciate the
hospitality of the Aspen Center for Physics, where this
collaboration began.


\begin{thebibliography}{10}
\bibitem{images} The flow field around each particle is modified by the presence
of the bottom boundary; within the potential flow approximation
having no mass flux across the plate  requires an "image" particle
beneath the plate, playing a similar role to image charges in
electrostatics. Our estimates indicate that these image particles
do not have a significant effect on the particle interactions.
\end{thebibliography}
\end{document}